\documentstyle[preprint,prl,aps]{revtex}
\begin{document}
\draft
\title{Vortex lattice melting in 2D superconductors and
Josephson arrays}

\author{M. Franz and S. Teitel\\}
\address{Department of Physics and Astronomy, University of Rochester,
Rochester, NY 14627\\}
\date{\today}
\maketitle

\begin{abstract}
Monte Carlo simulations of 2D vortex lattice melting
in a thin superconducting film (or alternatively an array
of Josephson junctions) are performed in the London limit.
Finite size scaling analyses are used to make a detailed
test of the dislocation mediated melting theory of KTNHY.
We find that the melting transition is weakly first order,
with a jump in shear modulus very close to that predicted
by the KTNHY theory.  No hexatic liquid phase is found.
\end{abstract}

\pacs{PACS number: 64.60-i, 74.60-w, 74.76-w}

\newpage
\narrowtext

Interest in the melting transition of
two dimensional (2D) vortex lattices has revived recently, primarily
due to the belief that the strongly fluctuating, layered, high-$T_c$
superconductors may behave 2D like in sufficiently large magnetic
fields \cite{ffh}.  This 2D melting transition has generally been
believed to be described by the Kosterlitz-Thouless-Nelson-Halperin-Young
(KTNHY) theory \cite{KTHNY,fisher,doniach} of dislocation mediated melting.
However, the very existence of a 2D vortex lattice at
any finite temperature has been recently questioned by Moore \cite{moore1},
who first argued that phase fluctuations destroy such a lattice,
and then found support for this picture from Monte Carlo (MC)
simulations \cite{moore2}. High order, high
temperature perturbative expansions \cite{brezin}, similarly show no evidence
for freezing into a vortex lattice in 2D.
Other MC simulations \cite{xing,hu}, however, find clear evidence for a
melting transition at finite temperature.  Hu and MacDonald \cite{hu} find
this transition to be first order, in opposition to the KTNHY prediction.

The above cited simulations, have all been performed in the
\lq\lq lowest Landau level" approximation, in which
the complex order parameter $\psi({\bf r})$ is expanded in terms of
eigenstates of the Gaussian part of the Landau-Ginsburg free-energy
functional.  Alternatively, one may use instead the London approximation,
in which
the amplitude of $\psi({\bf r})$ is assumed to be constant, and only
the phase is allowed to vary.  In this limit, the problem can be
efficiently simulated by utilizing the well known mapping \cite{minnhag}
onto the 2D Coulomb gas.  Logarithmically interacting
point charges model vortices in the phase of $\psi({\bf r})$.
For a uniform magnetic
field $B$, one has a fixed density $B/\Phi_0$ of positive integer charges,
on a uniform neutralizing background ($\Phi_0$ is the flux quantum).
The London approximation should be valid whenever the bare vortex core radius
$\xi_0 \ll a_v$, the average spacing between vortices.
This corresponds to temperatures well below the mean field transition
temperature,
where vortex lattice melting is expected to occur.
Earlier simulations of this 2D Coulomb gas, in the context of the
2D one component plasma problem \cite{ocp}, show clear evidence
for a finite temperature melting transition, and suggest that the transition
is weakly first order.
In the following, we report on new simulations in this
London approximation, in which we carry out the first finite
size scaling analysis of the melting transition, making a detailed
comparison with the KTNHY theory.  We show that the shear modulus
jumps discontinuously to zero at the melting transition with a value
very close to the universal KTNHY prediction, however we find
no evidence for a hexatic liquid phase.  We perform the
first conclusive test of the order of the melting transition
(within the London approximation) using the
histogram method.  We find that the transition is weakly
first order, consistent with earlier suggestions.

The model we simulate is given by the Hamiltonian,
\begin{equation}
{\cal H} = {1\over 2} \sum_{ij} (n_i-f)(n_j-f)V({\bf r}_i -{\bf r}_j).
\label{e1}
\end{equation}
In order to reduce the size of phase space, we have discretized the
continuum by
constraining the charges to the sites $i$ of a periodic triangular grid with
spacing $a_0$.  The sum is over all pairs of sites in an $L\times L$
parallelogram-shaped cluster of triangular grid, and $n_i=0$, or $+1$, is
the point charge (vorticity) at site $i$.  The neutralizing background
charge is
$f=(\sqrt 3 a_0^2/2)(B/\Phi_0)=(a_0/a_v)^2$, and $V({\bf r})$ is
the lattice Coulomb potential in 2D, which solves,
\begin{equation}
\Delta^2 V({\bf r}) = -2\pi \delta_{\bf r,0}
\label{e2}
\end{equation}
subject to periodic boundary conditions. Here $\Delta^2$ is the discrete
Laplacian.  To keep the energy finite, it is necessary to preserve
total charge
neutrality, which leads to the constraint $N_c\equiv\sum_i n_i$
$=Nf$, where $N=L^2$ is the total number of sites in the grid,
and $N_c$ is the total number of charges in the system.
Thus $f$ is the density of charges.
Further details may be found in Ref. \cite{lee}.
The connection
between Eq.\ (\ref{e1}) and the superconducting system is obtained by
measuring
the temperature of the Coulomb gas model, in units of
$\Phi_0^2 d/8\pi^2\lambda^2$,
where $\lambda$ is the magnetic penetration length, and $d$ the thickness,
of the
superconducting layer \cite{fisher,minnhag}.

The Hamiltonian (\ref{e1}) has been studied extensively
\cite{lee,lee2} in the $dense$ charge limit ($f=1/2,1/3$). Here
we are interested in the $dilute$ limit $f \ll 1$, where we expect
our discretized model to well approximate the continuum
(see Refs. \cite{ocp} for similar simulations, directly in the continuum).
We always choose $f$ commensurate with the system length $L$,
so that the ground state will be a perfect triangular charge
(vortex) lattice.  We study various charge densities
$f=1/m^2$, with $m=3$ to $12$, and fixed $N_c\approx 100$.
Detailed finite size scaling analyses are carried out for the specific case
of $f=1/49$, and $N_c=16,25,\dots,169$.

Our MC updating scheme is as follows.
In each MC step one charge is selected at random and moved
to a different site within a radius $\sim a_v/2$. This excitation
is then accepted or rejected according to the standard Metropolis algorithm.
$N_c$ such attempts we refer to as one MC sweep.
At low temperature, we also make global moves, by attempting to shift
entire rows of charges by one space.  Such moves are meant to model long
wavelength shear excitations, and help to accelerate equilibration near
the vortex lattice melting transition.  Data is collected by
heating the system up from the ground state. At each temperature
we discard $30,000$ MC sweeps to equilibrate the system. Then, starting from
this equilibrated configuration, we perform several (typically $4-6$)
independent
runs of $100,000$ sweeps each to sample physical quantities.  Errors are
estimated
from the standard deviation of these independent runs.
To verify consistency of our results, we also perform cooling from a random
configuration at high temperature; no substantial hysteresis is found.

The physical quantities we measure are: ($i$) the inverse dielectric
function,
\begin{equation}
\epsilon^{-1}(T)\equiv\lim_{k\rightarrow 0}\biggl\{1-{2\pi\over k^2TN}
                 \langle n_{\bf k} n_{-\bf k} \rangle \biggr\},
\label{e3}
\end{equation}
where $n_{\bf k}=\sum_i n_i\exp(-i {\bf k\cdot r}_i)$.
The vanishing of $\epsilon^{-1}$ upon heating signals
an ``insulator-conductor" transition in the Coulomb gas. As $\epsilon^{-1}$
can be mapped onto the helicity modulus of the superconductor \cite{XY-cgs},
its vanishing
signals the loss of superconducting phase coherence.
In the simulation, the ${\bf k}\rightarrow 0$ limit is
approximated by averaging $\epsilon^{-1}$ over the three smallest allowed
wave vectors; ($ii$) the six-fold orientational order correlation,
\begin{equation}
\varphi_6(T)\equiv {1\over N^2_c}\sum_{ij}\biggl\langle
              e^{i6(\theta_i-\theta_j)}\biggr\rangle
\label{e4}
\end{equation}
where the sum is over sites with non-vanishing charges $n_i=+1$, and
$\theta_i$ is the angle of the bond from $n_i$ to its nearest neighbor,
relative to some fixed reference direction; and ($iii$) the structure
function
\begin{equation}
S({\bf k})\equiv {1\over N_c} \langle n_{\bf k} n_{-\bf k} \rangle
           ={1\over N_c}\sum_{ij}e^{i{\bf k}\cdot ({\bf r}_i-{\bf r}_j)}
           \langle n_{i} n_{j} \rangle.
\label{e5}
\end{equation}

In Fig.\ 1, we plot $\epsilon^{-1}(T)$
and $\varphi_6(T)$, versus $T$, for $f=1/49$ and $N_c=169$.
The behavior of $\varphi_6(T)$ indicates two separate transitions at
$T_c(f)$ and $T_m$.  For a simple visualization of the resulting
three phases, we show in Fig.\ 2 intensity plots of $S({\bf k})$,
for {\bf k}'s in the first Brillouin zone (BZ).  We show results for
three different temperatures, with the data for each value of $T$
restricted to one third of the BZ.
For $T=0.003$ (Fig.\ 2a),
just below $T_c(f)$, we see a regular array of $\delta$-function Bragg peaks,
indicating long range translational order.
Thus for $T<T_c(f)$, we have a vortex lattice which is pinned to the
discretizing grid.
For $T=0.0065$ (Fig.\ 2b), just below $T_m$, we see a regular array of peaks,
but the peaks are now of finite width.  These peaks are consistent with
power law singularities, characteristic of the algebraic translational
order expected for a 2D lattice in the continuum.
Thus for $T_c(f)<T<T_m$, we have a ``floating" vortex lattice,
which is depinned from the grid, and we have reached the continuum limit.
For $T=0.0075$ (Fig.\ 2c), slightly above $T_m$, we see a rotationally
invariant structure ($\varphi_6\sim 0$),
typical for a liquid with short range correlations.
Thus for $T>T_m$, the floating lattice has melted into a liquid.

Returning to Fig.\ 1, we see that $\epsilon^{-1}$ vanishes at the depinning
transition $T_c(f)$.  Thus the floating lattice has lost superconducting
phase coherence.  This is nothing more than a reflection of the flux flow
resistance to be expected from an unpinned vortex lattice, which is free
to drift transversly to an applied d.c. current.  Our results explicitly show
that the absence of phase coherence
in this ${\bf k}\to 0$ sense, does not
not imply the absence of a well defined vortex lattice.

In the inset to Fig.\ 1, we show the dependence of $T_c(f)$ and $T_m$ on the
charge density $f$.  We see that only for sufficiently dilute systems,
$f<1/25$,
is there a floating lattice phase; for $f>1/25$ there is only a single
transition from a pinned lattice to a liquid.  As $f$ decreases, $T_c(f)\to 0$
as $\sim f$, while $T_m$ quickly approaches a finite constant
$T_m=0.0070\pm0.0005$, in good agreement with the melting temperature found
in earlier continuum simulations \cite{ocp}.
In terms of the superconductor, this means a vortex
lattice melting at $T_m=0.0070\>\Phi_0^2d/8\pi^2\lambda^2$,
well within the bounds estimated by Fisher
\cite{fisher} from the KTNHY theory.

The transition at $T_c(f)$ is an artifact of our discretization of
the continuum, and hence has no direct physical meaning for a uniform
continuous
superconductor.  However $T_c(f)$ does represent a physical depinning
transition
for the related problem of vortex states in periodic superconducting networks,
such as Josephson junction arrays.  In this case, our result that $T_c(f\to 0)$
vanishes, is consistent with early commensurability
arguments by Teitel and Jayaprakash \cite{tj}.
The result that a floating lattice exists above $T_c(f)$ is, however, a new
observation in the Josephson array context; the melting of this lattice at a
finite $T_m$ may dominate the physics of such arrays at small $f$.

To investigate the nature of the melting transition $T_m$, we have carried out
detailed finite size scaling analyses for the case $f=1/49$, in which $T_m$
is well separated from $T_c$.  Our approach is guided by
the KTNHY theory \cite{KTHNY}.
For a 2D lattice in the continuum, translational correlations
decay algebraically with a temperature dependent exponent,
$\langle e^{i{\bf G}\cdot({\bf r_i}-{\bf r_j})}\rangle
\sim |{\bf r_i}-{\bf r_j}|^{-\eta_{\bf G}(T)}$, where ${\bf G}$ is a reciprocal
lattice vector of the real space charge lattice.
For the 2D superconducting case, where the vortex compressibility is infinite,
$\eta_{\bf G}(T)=k_BT|{\bf G}|^2/4\pi\mu$, where $\mu$ is the vortex shear
modulus.
If ${\bf G}_1$ is the shortest reciprocal lattice vector, then
the KTNHY theory predicts that at $T_m$, $\eta_{{\bf G}_1}$ takes a
discontinuous
jump to infinity from the universal value of $\eta_{{\bf G}_1}(T_m^-)=1/3$.

To test this prediction for translational order, we measure the
height of peaks in the structure function.
{}From Eq.\ (\ref{e5}) we find that these should scale as
%
\begin{equation}
S({\bf G})\sim L^{2-\eta_{\bf G}(T)}\qquad\qquad {\rm for}\hskip 10pt T<T_m.
\label{e6a}
\end{equation}
Above $T_m$, translational order has exponential decay with a
correlation length $\xi$. One then obtains
\begin{equation}
S({\bf G}) \sim \xi^2\qquad\qquad {\rm for} \hskip 10pt T>T_m.
\label{e6b}
\end{equation}
In the Fig.\ 3a we plot $S({\bf G}_1)/L^2$,
as a function of $L$ on a log-log scale, for several
different temperatures.  Data for each temperature fall on
a straight line, confirming the expected power-law behavior.
These straight lines fall into three distinct groups.
For $T<T_c\simeq 0.0045$,
$S({\bf G}_1)/L^2\sim 1$, indicating the long range order of the pinned
lattice.
For $T_c<T<T_m\simeq 0.007$, we find algebraic decay,
$S({\bf G}_1)/L^2\sim L^{-\eta_{\bf G}(T)}$.  For $T>T_m$, we find
$S({\bf G}_1)/L^2\sim L^{-x}$, with $x\to 2$ as $T$ increases,
consistent with the short range order of a liquid.
The lines in Fig.\ 3a are a fit to Eq.(\ref{e6a}); the resulting
exponents $\eta_{{\bf G}_1}(T)$ are shown in Table 1.  We see that
$\eta_{{\bf G}_1}$ first exceeds the KTNHY universal value of $1/3$
at $T= 0.0065$, very close to our estimated melting transition of $T_m
\simeq 0.007$, where the slopes of
the lines in Fig.\ 3a show an apparent discontinuous jump.
Similar results, within the
``lowest Landau level" approximation, have very recently been obtained by
\v{S}\'{a}\v{s}ik and Stroud\cite{sasik}.

As a consistency check, we have also computed $S({\bf G_2})$, where
${\bf G}_2=2{\bf G}_1$.
Using similar fits as in Fig.\ 3a, we determine the exponent
$\eta_{{\bf G}_2}$, and show the results in Table. 1.
We see that $\eta_{{\bf G}_2}\simeq 4\eta_{{\bf G}_1}$ as expected,
since $\eta_{\bf G}\sim |{\bf G}|^2$.

We now consider the orientational order.
Below $T_m$, KTNHY predict long range $6-$fold orientational
order given by $\langle e^{i6(\theta(r)-\theta(0))}\rangle\sim$
$\alpha e^{-r/\xi_6} +\varphi_6^\infty$. For $\xi_6\ll L$, one obtains from
Eq.(\ref{e4})
\begin{equation}
\varphi_6\sim 2\pi\alpha\biggl({\xi_6\over L}\biggl)^2 + \varphi_6^\infty.
\label{e7}
\end{equation}
Above $T_m$, KTNHY predict a hexatic liquid phase, with algebraic orientational
order $\langle e^{i6(\theta(r)-\theta(0))}
\rangle\sim r^{-\eta_6(T)}$ with $\eta_6(T)<1/4$.  In such a case, one would
have,
\begin{equation}
\varphi_6\sim L^{-\eta_6}.
\label{e7'}
\end{equation}
At higher temperature, KTNHY predict a transition from the hexatic to an
isotropic liquid, with short ranged orientational order.
In this case, Eq.(\ref{e7}) again holds, but with
$\varphi_6^\infty = 0$ \cite{note2}.
In Fig.\ 3b we display $\varphi_6(T)$ as a function of
system size  $L$ for various temperatures. In the floating solid phase below
$T_m$ one clearly sees saturation of $\varphi_6(T)$ to a finite value as
$L$ increases. Solid lines represent a least square fits to Eq.\ (\ref{e7}),
and the extracted values of $\varphi_6^\infty$ are shown in Table 1. Above
$T_m$, we try fits to both Eqs.\ (\ref{e7}) and (\ref{e7'}).  We always find
that Eq.\ (\ref{e7}) gives the superior fit, and for $T>T_m$ we find,
within our numerical precision, $\varphi _6^\infty = 0$.

Thus our results for the floating lattice phase are consistent with
expectations for a 2D continuum lattice, and we find that translational
correlations at the melting transition are consistent with the KTNHY
prediction.  However we do not find any evidence for a hexatic liquid
above $T_m$.

The absence of the hexatic liquid suggests
the possibility that the melting transition
is not of the KTNHY type, but is perhaps weakly first order as found by
Hu and MacDonald \cite{hu}, and as suggested in Refs. \cite{ocp}.
To examine this possibility,
we measured the energy distribution $P(E)\sim e^{-F(E)/T}$ at the
melting temperature $T_m$
\cite{kosterlitz}, and in Fig.\ 4 we plot the resulting
free energy $F(E)$ versus  $E$. We see a double well structure
with an energy barrier $\Delta F$ between two coexisting phases.
In the inset to Fig.\ 4 we plot the dependence of $\Delta F$ on
system length $L$.  The growth in $\Delta F$ as $L$ increases is a
clear signal that the transition is first order, although
our sizes remain too small, and our data too noisy, to see
clearly the predicted scaling $\Delta F\sim L$.  To determine the
distributions in Fig.\ 4, we have computed $P(E)$ at fixed $T\simeq T_m$,
and then extapolated\cite{Swendsen} to determine $P(E)$ at
nearby $T$, finding the precise value of $T$ that gives
equal minima in $F(E)$.  In this way we obtain an improved estimate
$T_m\simeq 0.0066$.

To conclude, our results demonstrate that there is a clear
finite temperature melting transition of the vortex lattice
in two dimensions, within the London approximation.  This transition
is first order, with melting directly into an isotropic vortex liquid; no
hexatic liquid is found.  The first order transition however is very weak, so
that the jump at melting in $\eta_G$ (and hence in the vortex lattice shear
modulus) remains very close to the KTNHY universal prediction.

The authors are grateful to T. Chen, D. A. Huse, D. R. Nelson, and
Z. Te\v{s}anovi\'{c} for useful discussions.
This work was supported by DOE grant DE-FG02-89ER14017.  One of us (M.F.)
acknowledges the Rush Rhees Fellowship of the University of Rochester for
support in the initial stages of this project.


\bibliographystyle{unsrt}

\begin{thebibliography}{99}
\bibitem{note2}  In principle, the discretizing mesh acts like an ordering
field
		for $6-$fold orientational order, and so even in the isotropic
		liquid one might find
		$\varphi_6^\infty$ small but finite.  The vanishing values of
		$\varphi_6^\infty$
		which we find numerically above $T_m$ indicate that this is a very small
		effect at the density $f$ which we are studying.
\end{thebibliography}

\newpage
\begin{center}
\begin{tabular}{|c||c|c|c|} \hline
  $T$  &  $\eta_{{\bf G}_1}(T) $ &  $\eta_{{\bf G}_2}(T) $ &
  $\varphi_6^\infty$   \\ \hline\hline
 0.00475 & 0.188 $\pm$0.008   & 0.704 $\pm$0.055  & 0.571 $\pm$0.007  \\
 0.00500 & 0.207 $\pm$0.007   & 0.806 $\pm$0.032  & 0.529 $\pm$0.005  \\
 0.00525 & 0.211 $\pm$0.007   & 0.852 $\pm$0.028  & 0.504 $\pm$0.004  \\
 0.00550 & 0.248 $\pm$0.005   & 0.998 $\pm$0.019  & 0.476 $\pm$0.003  \\
 0.00575 & 0.255 $\pm$0.008   & 0.999 $\pm$0.029  & 0.458 $\pm$0.003  \\
 0.00600 & 0.296 $\pm$0.006   & 1.065 $\pm$0.028  & 0.426 $\pm$0.007  \\
 0.00625 & 0.319 $\pm$0.010   & 1.191 $\pm$0.016  & 0.403 $\pm$0.004  \\
 0.00650 & 0.4   $\pm$0.16    & 1.4   $\pm$0.22   & 0.33  $\pm$0.030  \\
 0.00675 & 1.4   $\pm$0.31    & 2.0   $\pm$0.31   & 0.20  $\pm$0.041  \\
 0.00750 & 3.4   $\pm$0.37    & 3.4   $\pm$0.44   & 0.03  $\pm$0.046   \\
 0.01100 & 2.8   $\pm$0.23    & 2.9   $\pm$0.30   &-0.01  $\pm$0.032  \\
 0.01500 & 2.2   $\pm$0.12    & 2.1   $\pm$0.22   & 0.00 $\pm$0.020  \\ \hline
\end{tabular}
\end{center}
\begin{description}
\item{Table 1:} Temperature dependence of the exponents $\eta_{{\bf G}_1}(T)$
and
$\eta_{{\bf G}_2}(T)$. Also displayed limiting values of $\varphi_6(T)$ for $L
\rightarrow \infty$, $\varphi_6^\infty$.
\end{description}

\begin{figure}
\end{figure}

\begin{description}
\item{Fig.1} Inverse dielectric function $\epsilon^{-1}(T)$ and
orientational order correlation $\varphi_6(T)$ versus $T$
for $f=1/49$ and $N_c=169$.  Inset shows the dependence of the depinning
and melting temperatures, $T_c$ and $T_m$, on charge density $f$.
Solid and dashed lines are guides to the eye only.

\item{Fig.2} Structure function $S({\bf k})$ in the first Brillouin zone,
(BZ) for $f=1/49$ and $N_c=63$, and three different temperatures $T$.
Data for each $T$ is restricted to one third of the BZ.
(a) $T=0.003$, just below $T_c$, in the
``pinned lattice" state. (b) $T=0.0065$, just below $T_m$, in the
``floating lattice" state. (c) $T=0.0075$, just above $T_m$, in the liquid.
Intensities are plotted nonlinearly to enhance features.

\item{Fig.3} (a) Finite size scaling of $S({{\bf G}_1})/L^2$ (note the
log-log scale).
Solid and dashed lines are fits to Eqs.\ (\ref{e6a}). (b)
Finite size scaling of $\varphi_6(T)$.  Solid and dashed lines are fits to
eq.\ (\ref{e7}).

\item{Fig.4} Free energy distribution $F(E)$ versus $E$, at melting $T_m$,
for several system sizes $L$.  The growth in energy barrier $\Delta F$ with
increasing $L$ (see inset) indicates a first order transition.  Curves
for different $L$ are offset from each other by a constant, for the sake
of clarity.

\end{description}

\end{document}